\documentclass[runningheads]{llncs}

\usepackage[T1]{fontenc}
\usepackage{graphicx}
\usepackage{url}
\usepackage{amsmath}
\usepackage{amssymb}
\usepackage{subcaption} 

\begin{document}

\title{Evaluating Asynchronous Semantics in Trace-Discovered Resilience Models:\\A Case Study on the OpenTelemetry Demo}

\titlerunning{Trace-Discovered Resilience Models with Async Semantics}

\author{Anatoly A.~Krasnovsky\inst{1,2}}
\authorrunning{Krasnovsky}

\institute{
    Innopolis University, Innopolis, Russia 
    \and
    MB3R Lab, Innopolis, Russia\\
    \email{a.a.krasnovsky@gmail.com}
}
\maketitle

\begin{abstract}
While distributed tracing and chaos engineering are becoming standard for microservices, resilience models remain largely manual and bespoke. We revisit a trace-discovered connectivity model that
derives a service dependency graph from traces and uses Monte Carlo simulation
to estimate endpoint availability under fail-stop service failures. Compared to
earlier work, we (i) derive the graph directly from raw OpenTelemetry traces,
(ii) attach endpoint-specific success predicates, and (iii) add a simple
asynchronous semantics that treats Kafka edges as non-blocking for immediate
HTTP success. We apply this model to the OpenTelemetry Demo (``Astronomy
Shop'') using a GitHub Actions workflow that discovers the graph, runs
simulations, and executes chaos experiments that randomly kill microservices in
a Docker Compose deployment. Across the studied failure fractions, the model
reproduces the overall availability degradation curve, while asynchronous
semantics for Kafka edges change predicted availabilities by at most about $10^{-5}$ (0.001 percentage points). This null result suggests that for immediate HTTP availability in this case study, explicitly modeling asynchronous dependencies is not warranted, and a simpler connectivity-only model is sufficient.
\keywords{microservices \and resilience \and chaos engineering \and availability modeling \and model discovery \and fault tolerance}
\end{abstract}

\section{Introduction}
\label{sec:introduction}

Microservice architectures are now a dominant way to build cloud and edge
applications, promising independent deployment, technology heterogeneity, and
faster evolution of large systems~\cite{dragoni2017microservices}. At the same
time, decomposing a system into dozens of services, databases, and message
brokers introduces a large space of failure modes: a single user request may
transit many components, and any subset of them may be unavailable or slow. For
operators and SRE teams, understanding how these structural dependencies affect
end-to-end availability is essential for setting realistic SLOs and making safe
decisions during incidents~\cite{billinton1992reliability,trivedi2016probability}.

Chaos engineering tackles this problem empirically by injecting controlled
faults (such as process crashes, latency, or network partitions) into
production-like environments and checking whether the system still meets its
SLOs~\cite{basiri2016chaos}. Recent reviews report growing adoption but also stress that large campaigns
remain operationally expensive and risky to run broadly and frequently~\cite{owotogbe2024chaosmlr};
complementary empirical work in event-driven microservices reports similar practical challenges~\cite{adapa2025eventchaos}.
As a result, organisations typically reserve fault injection for a relatively small set of scenarios
and environments.

In parallel, distributed tracing has become a standard observability signal.
Systems such as Dapper, X-Trace and Pivot
Tracing~\cite{sigelman2010dapper,fonseca2007xtrace,mace2015pivottracing} and
the OpenTelemetry/Jaeger stack~\cite{opentelemetryTraces,jaeger2017} record
per-request spans that expose causal service-dependency graphs, which
practitioners already use for debugging and impact
analysis~\cite{sambasivan2014tracinginsights}. Our recent ``model discovery''
work~\cite{krasnovsky2025modeldiscovery} showed that a simple connectivity-only
service graph mined from Jaeger's \texttt{/api/dependencies} endpoint, plus
replica counts, suffices to approximate endpoint availability on the
DeathStarBench Social Network benchmark under fail-stop
faults~\cite{gan2019deathstarbench}.

This paper revisits that idea on the OpenTelemetry Demo (``Astronomy
Shop'')~\cite{opentelemetryDemoDocs,opentelemetryDemoRepo} and refines the
model along two axes. First, we derive the service graph directly from raw
OpenTelemetry traces, exposing per-span attributes that let us identify
Kafka-based edges as asynchronous. Second, we introduce endpoint-specific
success predicates and an optional asynchronous semantics in which Kafka edges
are treated as non-blocking for immediate success of selected HTTP endpoints.
Since our success predicates are defined on the immediate HTTP response (not eventual processing), async semantics can matter
only when Kafka edges lie on paths to required synchronous targets; we test this empirically in Sect.~\ref{sec:results}.
A GitHub Actions workflow discovers the graph, runs Monte Carlo simulations and
executes chaos experiments that randomly kill microservices in a Docker Compose
deployment. The contributions are: (i) a refinement of trace-discovered
resilience models with trace-level graph extraction, endpoint predicates and
asynchronous edges; (ii) a CI workflow for the OpenTelemetry Demo that
combines trace collection, model discovery, Monte Carlo simulation and chaos
experiments; and (iii) an empirical negative result: for the studied immediate HTTP SLOs, treating Kafka edges as non-blocking changes predicted availability by $<10^{-5}$, giving a criterion for when connectivity-only models suffice.

\section{Background and Related Work}
\label{sec:background}

Microservices improve agility~\cite{dragoni2017microservices} but complicate failure reasoning due to interacting faults~\cite{gan2019deathstarbench}. While classical reliability engineering uses Monte Carlo methods to estimate availability on probabilistic graphs~\cite{billinton1992reliability,trivedi2016probability}, chaos engineering injects faults in production to verify SLOs empirically~\cite{basiri2016chaos}. However, broad chaos campaigns remain costly~\cite{owotogbe2024chaosmlr,adapa2025eventchaos}.
In parallel, distributed tracing systems (e.g., OpenTelemetry/Jaeger~\cite{opentelemetryTraces,jaeger2017}) expose causal service graphs~\cite{sambasivan2014tracinginsights}. Our previous work~\cite{krasnovsky2025modeldiscovery} showed that connectivity-only graphs mined from Jaeger plus replica counts suffice to approximate availability on DeathStarBench~\cite{gan2019deathstarbench}. This paper extends that approach with raw trace extraction, endpoint predicates, and asynchronous semantics.

\section{Trace-Discovered Resilience Model with Async Semantics}
\label{sec:model}

We now summarise the trace-discovered resilience model used in this paper. The
model starts from a dependency graph derived from distributed traces, defines a
fail-stop failure model over services, and estimates endpoint availability via
Monte Carlo simulation. Compared to~\cite{krasnovsky2025modeldiscovery}, we add
endpoint-specific success predicates and explicit asynchronous edges.

\subsection{Service Dependency Graph from Traces}
\label{sec:graph}

Let $S$ be the set of application services and let $E \subseteq S \times S$ be a
directed service-dependency graph. We obtain $E$ from traces collected by
OpenTelemetry and exported to Jaeger~\cite{opentelemetryTraces,jaeger2017}.
Each span carries a service name and protocol-specific attributes (HTTP, RPC or
messaging). By aggregating parent--child relationships and projecting to the
service level, we add an edge $(s_i, s_j) \in E$ whenever service $s_i$ calls or
sends a message that is later consumed by service $s_j$. Concretely, we parse each exported trace as a span tree, map spans to services via the resource attribute
\texttt{service.name}, and reconstruct parent--child links using \texttt{span\_id} and \texttt{parent\_span\_id}. We add an edge
from the parent span’s service to the child span’s service and then deduplicate edges on $(s_i,s_j)$ across all traces. We filter
out infrastructure spans and very short-lived internal instrumentation before aggregation.

In the OpenTelemetry Demo artifact~\cite{krasnovsky2025oteldemoresilience},
scripted queries to Jaeger export recent traces to JSON, which we convert into a
graph description \texttt{graph.json}. The graph lists the service set $S$, the
edge set $E$, and a flag for edges that traverse Kafka, detected via messaging
semantic conventions in span attributes~\cite{opentelemetryTraces}. We
optionally associate each service $s \in S$ with a replication factor
$r(s) \in \mathbb{N}$; when explicit replication data are absent we assume
$r(s) = 1$.

For the deployment studied here, the resulting graphs are small but
non-trivial: each run yields $|S| = 16$ services in the graph, of which
$15$ microservices are eligible for failure injection
(Sect.~\ref{sec:failure-model}). The number of directed edges varies slightly
across runs, between $22$ and $30$ (median $|E| = 23$). Exactly three edges
are tagged as Kafka-based asynchronous edges, namely
\texttt{checkout} $\rightarrow$ \texttt{kafka},
\texttt{kafka} $\rightarrow$ \texttt{accounting}, and
\texttt{kafka} $\rightarrow$ \texttt{fraud-detection}, so that
$|A| = 3$ (about $13\%$ of edges in a typical run).

Instead of maintaining a manually curated ``ground truth'' dependency graph, we treat the trace-derived graph as an operational
representation of the running system. Documentation and \texttt{docker-compose.yml} capture intended interactions but may miss
dynamic behaviour (retries, fallbacks, background jobs), and keeping a manual graph in sync undermines automation. We therefore
apply lightweight sanity checks (e.g., deployed vertices, no unknown targets) and use the live chaos experiments
(Sect.~\ref{sec:results}) as end-to-end validation.

\subsection{Failure Model and Monte Carlo Evaluation}
\label{sec:failure-model}

Let $S_{\mathrm{elig}} \subseteq S$ be the set of services that are eligible to be
killed in experiments (excluding, for example, tracing backends and the load
generator). For a given failure fraction $p_{\mathrm{fail}} \in (0,1)$ we model a failure
scenario as a subset $K \subseteq S_{\mathrm{elig}}$ of failed services. In the
implementation we sample $|K| \approx p_{\mathrm{fail}} \cdot |S_{\mathrm{elig}}|$ services
uniformly without replacement, mirroring the behaviour of the chaos
harness~\cite{krasnovsky2025oteldemoresilience}. The corresponding alive service
set is $S_{\mathrm{alive}} = S \setminus K$.

Let $U$ denote the set of HTTP endpoints of interest. For each endpoint
$u \in U$ and model semantics $m$ (all-blocking or async; see
Sect.~\ref{sec:async-semantics}) we define a Boolean predicate
$\textit{success}^{(m)}(u, K)$ indicating whether $u$ is considered available
under failure scenario $K$. Given a failure probability $p_{\mathrm{fail}}$ we estimate the
availability of $u$ under semantics $m$ by Monte Carlo simulation:
\begin{equation}
  \hat{R}^{(m)}_u(p_{\mathrm{fail}}) =
  \frac{1}{M} \sum_{i=1}^{M} \textit{success}^{(m)}(u, K_i),
\end{equation}
where $K_1, \dots, K_M$ are independent samples of failure sets at fraction $p_{\mathrm{fail}}$. This is the usual Monte Carlo estimator for network reliability~\cite{billinton1992reliability,trivedi2016probability}. In the CI workflow each job uses a fixed sample size of $M = 5 \times 10^{6}$ Monte Carlo trials. For a single endpoint with true availability $R$, the binomial standard error of the estimator is
\[
  \mathrm{SE}_\mathrm{MC} = \sqrt{R(1-R)/M}.
\]
In the worst case $R = 0.5$ this gives $\mathrm{SE}_\mathrm{MC} \le 2.3 \times 10^{-4}$ (about 0.023 percentage points). Section~\ref{sec:monte-carlo-config} further averages these estimates over 50 jobs per configuration, so the Monte Carlo contribution to the discrepancies is negligible.

Algorithmically, each Monte Carlo trial: (1) samples a failure set $K \subseteq S_{\mathrm{elig}}$;
(2) forms the alive graph by removing $K$ (and, for async semantics, edges $A$);
(3) runs BFS from $e(u)$ to find reachable services; and (4) evaluates $\textit{success}^{(m)}(u, K)$.

\subsection{Endpoint Success Predicates}
\label{sec:model:endpoints}

To relate graph-level reachability to client-perceived success we define, for
each endpoint $u \in U$:
\begin{itemize}
  \item an entry service $e(u) \in S$, which receives the HTTP request; and
  \item a finite set of target services $T(u) \subseteq S$ with a combinator
        that encodes the success rule.
\end{itemize}
The success rule is one of the following:

\textbf{all\_of} ($u$ succeeds if all services in $T(u)$ are reachable from $e(u)$ in the alive graph under semantics $m$);

\textbf{any\_of} ($u$ succeeds if at least one service in $T(u)$ is reachable); or

\textbf{k\_of\_n} ($u$ succeeds if at least $k$ out of $|T(u)|$ services in $T(u)$ are reachable).
Endpoint configurations are stored in a JSON file
(\texttt{config/targets.json}) that maps HTTP routes to entry services, target sets and combinators. 

\begin{table}[t]
  \centering
  \caption{Endpoint success predicates used in the experiments. The \emph{Targets}
  column lists the backend services that must be reachable from the entry service
  $e(u)$ under the stated rule.}
  \label{tab:endpoints}
  \begin{tabular}{llll}
    \hline
    Endpoint $u$ & Entry $e(u)$ & Targets $T(u)$ & Rule \\
    \hline
    \texttt{GET /api/products} & \texttt{frontend} & \{\texttt{product-catalog}\} & \emph{all\_of} \\
    \texttt{GET /api/recommendations} & \texttt{frontend} & \{\texttt{recommendation}\} & \emph{all\_of} \\
    \texttt{GET /api/cart} & \texttt{frontend} & \{\texttt{cart}\} & \emph{all\_of} \\
    \texttt{POST /api/checkout} & \texttt{frontend} & \{\texttt{checkout}, \texttt{cart}, \texttt{payment}, \texttt{shipping}\} & \emph{all\_of} \\
    \hline
  \end{tabular}
\end{table}

For the four endpoints we actively probe in this study, the predicates are detailed in Table~\ref{tab:endpoints}. Under these definitions, immediate HTTP success depends only on synchronous backend services; Kafka consumer services appear in the discovered graph but are not in any $T(u)$, so removing async edges can change predictions only when they lie on paths to required synchronous targets.

\subsection{Asynchronous Edges and Semantics}
\label{sec:async-semantics}

Many microservices use brokers such as Kafka to decouple producers and consumers.
For user-facing SLOs defined over the \emph{immediate} HTTP response, downstream consumers may process
events later (or even fail without failing the HTTP request), so treating every observed dependency as
blocking can overestimate the impact of consumer outages on user-visible availability.

We capture this distinction by identifying a subset $A \subseteq E$ of
\emph{asynchronous} edges. In the OpenTelemetry Demo,
\texttt{traces\_to\_deps.py} tags edges as async when they correspond to spans
with \texttt{messaging.system = "kafka"} and producer/consumer span
kinds~\cite{opentelemetryTraces,krasnovsky2025oteldemoresilience}. All other
edges, originating from HTTP or RPC spans, are treated as synchronous.

Over the same graph and failure model we consider two semantics:
\begin{itemize}
  \item \textbf{All-blocking semantics.} All edges in $E$ are treated as required
        dependencies. Reachability is computed on the full alive graph, as
        in~\cite{krasnovsky2025modeldiscovery}.
  \item \textbf{Async semantics.} We treat edges in $A$ as non-blocking for immediate endpoint success. Operationally, we compute
        reachability on the alive graph with $A$ removed, so Kafka branches do not gate success when core synchronous dependencies
        remain reachable. This is an intentionally simple approximation scoped to Table~\ref{tab:endpoints}; it does not model broker
        acknowledgement modes or eventual-processing guarantees.
\end{itemize}

In general, asynchronous edges in $A$ can influence predictions only when
some required targets in $T(u)$ are reachable from $e(u)$ solely via those
edges. In our configuration (Table~\ref{tab:endpoints}), the probed endpoints
define immediate HTTP success entirely in terms of a small set of synchronous
backend services; the Kafka consumer services are not required members of any
$T(u)$. Under all-blocking semantics this implies that Kafka branches can only
matter if they lie on paths to those synchronous targets. Under async semantics
we remove edges in $A$ before computing reachability, so Kafka branches do not
cause an endpoint to be marked unavailable when its core synchronous
dependencies remain reachable.

\section{Experimental Methodology}
\label{sec:methodology}

We now describe how we instantiate the model for the OpenTelemetry Demo, design
the chaos experiments, and compare model predictions with live availability.

\subsection{Research Questions}

We evaluate: (1) the accuracy of blocking semantics on the demo; (2) whether async semantics reduce bias without harming accuracy; and (3) qualitative consistency with prior work~\cite{krasnovsky2025modeldiscovery}.

\subsection{System Under Test and Failure Model}

We deploy the OpenTelemetry Astronomy Shop demo in Docker Compose, following
the official documentation~\cite{opentelemetryDemoDocs,opentelemetryDemoRepo}.
From the running containers we designate a subset $S_{\mathrm{elig}}$ of
microservices as eligible for failure injection, excluding infrastructure such as
the tracing backend, the OpenTelemetry
Collector~\cite{opentelemetryCollectorResiliency}, and the synthetic load
generator. In our configuration, eligibility is determined by
\texttt{config/services\_disallowlist.txt} in the artifact~\cite{krasnovsky2025oteldemoresilience};
among the application services present in the graph this excludes only the
\texttt{frontend} entrypoint, while infrastructure services (such as Jaeger and
Prometheus) never appear in $S$ at all.

We consider failure fractions $p_{\mathrm{fail}} \in \{0.1, 0.3, 0.5, 0.7, 0.9\}$. For each
value we run $n = 50$ independent repetitions (“chunks”). Each chunk consists of
100 sequential chaos windows of length 60 s, giving 5000 windows per failure
fraction overall. In every chaos window the harness samples a set
$K \subseteq S_{\mathrm{elig}}$ of size $\approx p_{\mathrm{fail}} \cdot |S_{\mathrm{elig}}|$, terminates the chosen
containers, waits 15 s for the system to stabilise, and then issues active HTTP
probes for the next 40 s of the window. At the end of the window all services are
restarted before the next failure set is sampled.

The demo’s built-in Locust-based generator creates background traffic, but we
compute availability solely from the explicit probes issued by the harness.

\subsection{Monte Carlo Configuration}
\label{sec:monte-carlo-config}

For every failure fraction and semantics $m \in \{\textit{all}, \textit{async}\}$ we run Monte Carlo
simulation using the discovered graph. We estimate $\hat{R}^{(m)}_u(p_{\mathrm{fail}})$ for each
endpoint $u \in U$ using the $M$ sampled failure sets from Sect.~\ref{sec:failure-model} and the
endpoint predicates from Sect.~\ref{sec:model:endpoints}. We then compute a probe-weighted aggregate
estimate by averaging the endpoint-specific estimates under the probe distribution.
We use the same Monte Carlo sample size and configuration as in Sect.~\ref{sec:failure-model}.

\subsection{HTTP Probing and Live Availability}
\label{sec:live-availability}

During each chaos window the probing script issues a fixed number of HTTP
requests (100 in our implementation), selecting endpoints from the same set $U$
as the model. For every probe we record the target endpoint, the HTTP status
code, and whether the request is considered a success. Following
SRE-style practice in our previous study~\cite{krasnovsky2025modeldiscovery},
5xx responses, timeouts and transport-level errors are treated as failures; 2xx and 3xx/4xx codes are counted as successes.

For endpoint $u$ and failure fraction $p_{\mathrm{fail}}$ we let
$Y_{u,p,1}, \dots, Y_{u,p,n_{u,p}}$ be the binary success indicators for all
probes of $u$. The empirical live availability $\hat{R}^{\mathrm{live}}_u(p_{\mathrm{fail}})$ is the fraction of successful probes. We estimate standard errors via binomial variance; for $5 \times 10^{5}$ probes, errors are typically $< 0.5\%$. We report 95\% confidence intervals and a probe-weighted aggregate availability $\hat{R}^{\mathrm{live}}(p_{\mathrm{fail}})$.

When interpreting differences between model predictions and live measurements we
treat absolute availability gaps smaller than
$\Delta_{\min} = 0.01$ (one percentage point) as practically negligible for this
case study. This threshold is several times larger than the live and Monte Carlo
standard errors but matches the typical granularity at which SREs reason about
changes in endpoint availability.

\subsection{Implementation and CI Pipeline}

All steps are automated via a GitHub Actions workflow in the repository~\cite{krasnovsky2025oteldemoresilience}.
For each $(p_{\mathrm{fail}}, \textit{chunk})$ combination the workflow:
\begin{enumerate}
  \item checks out the repository and starts the OpenTelemetry Demo in Docker
        Compose;
  \item warms up the system and collects recent traces from Jaeger;
  \item constructs the dependency graph and Kafka edge annotations;
  \item runs Monte Carlo simulations for both semantics on the discovered graph;
  \item executes the chaos harness and HTTP probing; and
  \item aggregates results and uploads JSON/CSV artefacts.
\end{enumerate}
Step~2 performs trace collection before any failures are injected. The helper
script waits for the demo to become responsive under its built-in
Locust-generated background load and then uses Jaeger's HTTP API to download
traces for all application services over a fixed recent time window. We rely on
the demo's default collector configuration, which records every request as a
trace, and do not introduce any additional sampling in our scripts. In
practice this yields on the order of hundreds to thousands of traces per job,
which are then collapsed into the service graph as described in
Sect.~\ref{sec:graph}.

\section{Results}
\label{sec:results}

We now compare the trace-discovered resilience model with the live chaos
experiments on the OpenTelemetry Demo.

\subsection{Aggregate Availability and Bias}

\begin{figure}[t]
  \centering
  \begin{subfigure}[b]{0.495\textwidth}
    \includegraphics[width=\linewidth]{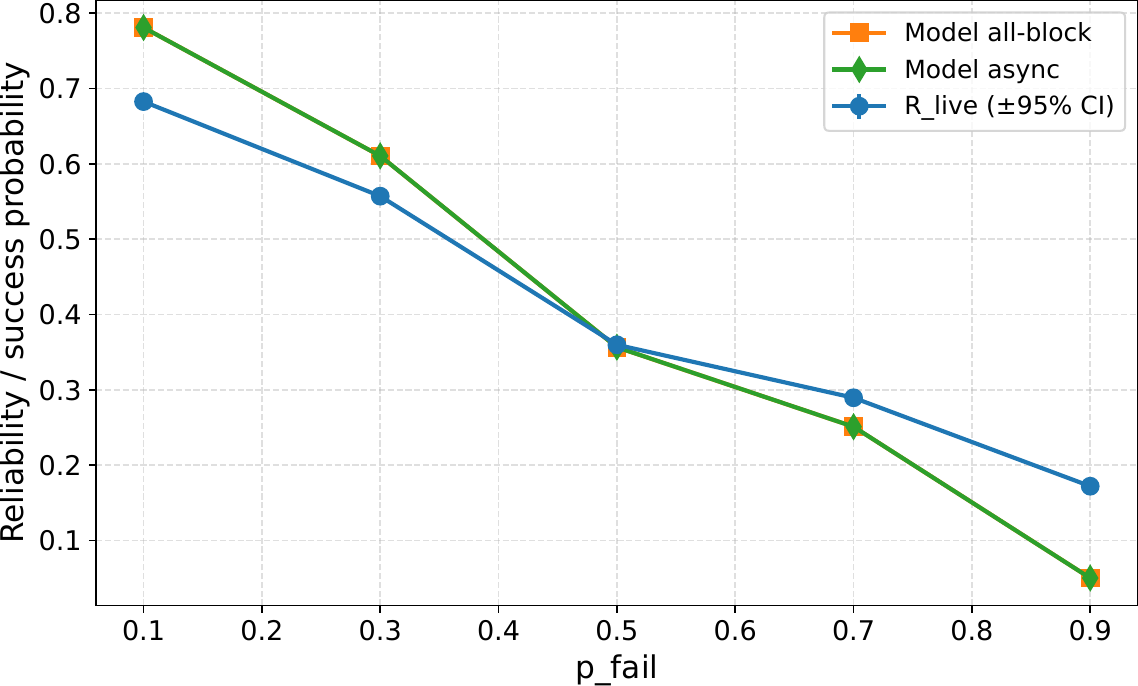}
    \caption{Aggregate availability}
    \label{fig:results:abs}
  \end{subfigure}
  \hfill
  \begin{subfigure}[b]{0.495\textwidth}
    \includegraphics[width=\linewidth]{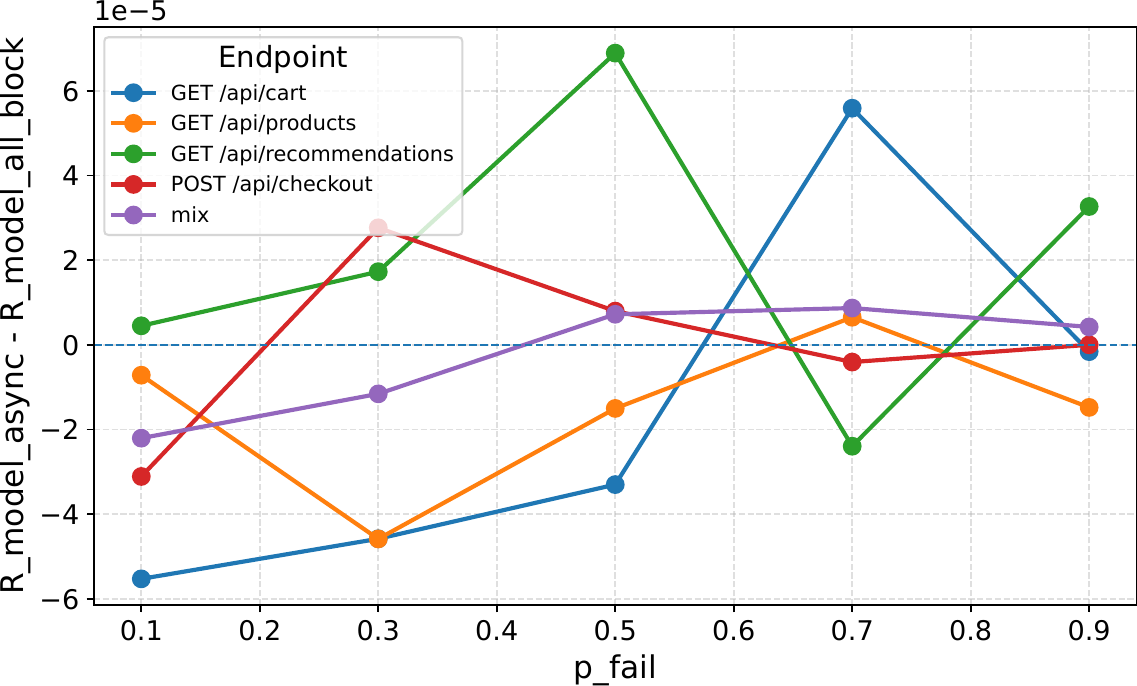}
    \caption{Async vs Blocking Delta}
    \label{fig:results:delta}
  \end{subfigure}
  \caption{Comparison of model predictions with live measurements. (a) Probe-weighted availability as a function of failure fraction $p_{\mathrm{fail}}$. (b) Endpoint-level difference $\Delta_u(p_{\mathrm{fail}})$ (Async minus All-blocking). Note the tiny $10^{-5}$ scale in (b).}
  \label{fig:results}
\end{figure}

Table~\ref{tab:global-summary} summarises, for each failure fraction
$p_{\mathrm{fail}}$, the probe-weighted aggregate live availability
$\hat{R}^{\mathrm{live}}(p_{\mathrm{fail}})$ and the model predictions under all-blocking and
async semantics. Figure~\ref{fig:results:abs} plots the same data with
95\% confidence intervals for the live measurements. Monte Carlo standard
errors are bounded by Sect.~\ref{sec:failure-model} to at most
$2.3 \times 10^{-4}$ (about $0.023$ percentage points) per endpoint, and are
further reduced by averaging over 50 jobs
(Sect.~\ref{sec:monte-carlo-config}). Error bars for the model curves would
therefore be smaller than the line thickness and are omitted from the plot
for visual clarity.

At low failure fractions ($p_{\mathrm{fail}} = 0.1$ and $0.3$) the model is
optimistic, overestimating live availability by roughly 10 and 5 percentage
points. At $p_{\mathrm{fail}} = 0.5$ the aggregate prediction is essentially
unbiased, while at higher failure fractions the model becomes pessimistic,
underestimating availability by about 4 points at $0.7$ and 12 points at $0.9$.
This pattern is similar to the DeathStarBench
study~\cite{krasnovsky2025modeldiscovery}: the connectivity-only model captures
the overall degradation curve but overestimates resilience when failures are
rare and underestimates it when a large fraction of services are down.

Notably, the all-blocking and async semantics are numerically identical at this aggregate level: the lines for $\hat{R}^{(\mathrm{all})}(p_{\mathrm{fail}})$ and $\hat{R}^{(\mathrm{async})}(p_{\mathrm{fail}})$ in Fig.~\ref{fig:results:abs} are visually indistinguishable.

\begin{table}[t]
  \centering
  \caption{Probe-weighted aggregate availability and bias by failure fraction
  $p_{\mathrm{fail}}$ (across endpoints). Each entry is estimated from 5000 chaos
  windows per failure fraction (50 independent chunks × 100 windows).}
  \label{tab:global-summary}
  \begin{tabular}{cccccc}
    \hline
    $p_{\mathrm{fail}}$ &
    $\hat{R}^{\mathrm{live}}$ &
    $\hat{R}^{(\mathrm{all})}$ &
    $\hat{R}^{(\mathrm{async})}$ &
    $\hat{R}^{(\mathrm{all})} - \hat{R}^{\mathrm{live}}$ &
    $\hat{R}^{(\mathrm{async})} - \hat{R}^{\mathrm{live}}$ \\
    \hline
    0.1 & 0.683 & 0.781 & 0.781 & +0.098 & +0.098 \\
    0.3 & 0.557 & 0.610 & 0.610 & +0.053 & +0.053 \\
    0.5 & 0.360 & 0.356 & 0.356 & $-0.003$ & $-0.003$ \\
    0.7 & 0.289 & 0.251 & 0.251 & $-0.038$ & $-0.038$ \\
    0.9 & 0.172 & 0.050 & 0.050 & $-0.122$ & $-0.122$ \\
    \hline
  \end{tabular}
\end{table}

\subsection{Distribution of Errors Across Conditions}

\begin{figure}[t]
  \centering
  \includegraphics[width=\linewidth]{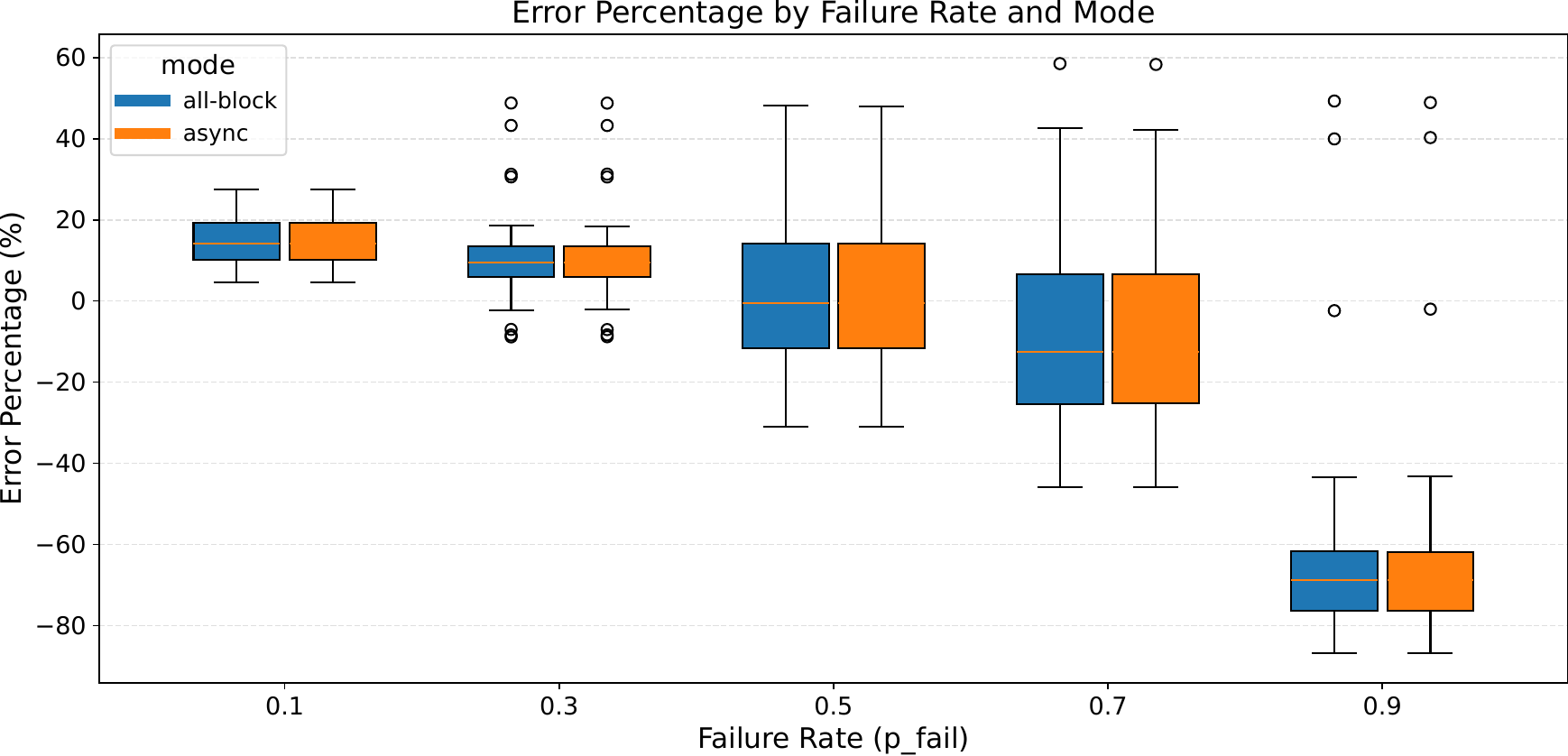}
  \caption{Distribution of relative error between model and live availability
  for each failure fraction $p_{\mathrm{fail}}$ and semantics. Boxes summarise
  per-endpoint percentage errors across repetitions; blue corresponds to the
  all-blocking semantics and orange to the async semantics.}
  \label{fig:error-boxplot}
\end{figure}

To understand per-endpoint behaviour we compute, for each endpoint $u$ and
failure fraction, the percentage error:

$\textit{err}^{(m)}_u(p_{\mathrm{fail}}) = 100 \times
(\hat{R}^{(m)}_u(p_{\mathrm{fail}}) - \hat{R}^{\mathrm{live}}_u(p_{\mathrm{fail}})) /
\hat{R}^{\mathrm{live}}_u(p_{\mathrm{fail}})$.\\
Figure~\ref{fig:error-boxplot} shows the distribution of these percentage errors
by $p_{\mathrm{fail}}$ and semantics.

At low failure fractions the distributions are concentrated around positive
values (modest optimism); at medium failure fractions the medians are close to
zero with a wider spread; and at high failure fractions they become strongly
negative, reflecting pessimism when most services are down. The per-endpoint
error distributions for all-blocking and async semantics almost completely
overlap. Mean per-endpoint absolute errors range from roughly 17\% to 28\%
across failure fractions, and the difference between the two semantics is well
within Monte Carlo noise for every configuration.

\subsection{Endpoint-Level Effect of Async Semantics}

To isolate the effect of asynchronous semantics we examine, for each endpoint
$u$ and failure fraction, the difference
$\Delta_u(p_{\mathrm{fail}}) = \hat{R}^{(\mathrm{async})}_u(p_{\mathrm{fail}}) -
\hat{R}^{(\mathrm{all})}_u(p_{\mathrm{fail}})$.
Figure~\ref{fig:results:delta} plots $\Delta_u(p_{\mathrm{fail}})$ for the four probed endpoints and for the probe-weighted mixture.

Across all endpoints and failure fractions the differences are extremely small
(on the order of $10^{-5}$) and oscillate around zero. This scale is roughly
three orders of magnitude below our practical negligibility threshold
$\Delta_{\min} = 0.01$ from Sect.~\ref{sec:live-availability} and more than an
order of magnitude smaller than the Monte Carlo standard error bounds in
Sect.~\ref{sec:failure-model}. Even for the checkout endpoint—whose flow
interacts most clearly with Kafka in the demo—the two semantics are therefore
indistinguishable at the resolution and power of our experiments. This is
consistent with the way we defined endpoint success predicates in
Sect.~\ref{sec:model:endpoints}: for this case study, immediate HTTP success is
determined solely by a small set of synchronous backend services, and the Kafka
consumer services are not required members of any $T(u)$. This null effect is informative for model selection: for endpoints whose SLO is defined on the immediate HTTP response and whose
required targets are synchronous, connectivity-only models can be used without loss at our experimental resolution. Modeling async
flows becomes justified mainly when success predicates encode eventual-processing guarantees (e.g., "order confirmed" implies
"shipping message consumed").

\subsection{Cross-Application Comparison}

Qualitatively, the OpenTelemetry Demo behaves similarly to the DeathStarBench
Social Network~\cite{krasnovsky2025modeldiscovery,gan2019deathstarbench}: a
connectivity-only graph with independent fail-stop faults captures the overall
availability degradation, with optimism at low failure fractions and pessimism
at high ones due to mechanisms such as retries and gray failures that the model
omits.

The main architectural difference is the Kafka-based event pipeline. Despite
this, treating Kafka edges as asynchronous changes predicted availabilities by
at most about $10^{-5}$, so for the endpoints and failure regimes we exercised,
immediate HTTP success is effectively determined by a small set of synchronous
backends. We view this null effect as demo-specific rather than evidence that
asynchronous dependencies never matter for availability.


\section{Threats to Validity}
\label{sec:threats}

Our results share common threats to validity for empirical studies of distributed
systems:

\textbf{Internal validity.} The model and harness share configuration but are separate implementations; 60 s windows and Docker health checks may miss delayed recovery or stateful failures.

\textbf{Construct validity.} Predicates approximate SLOs; trace-to-graph extraction may miss edges under sampling and ignores latency/gray failures.

\textbf{Conclusion validity.} We interpret effect sizes rather than formal tests; large sample sizes make statistical noise negligible compared to systematic bias.

\textbf{External validity.} The demo is a small benchmark; correlated faults or gray failures may yield different results.


\section{Conclusion and Future Work}
\label{sec:conclusion}

We extended a trace-discovered resilience modelling framework with
endpoint-specific success predicates and asynchronous semantics, and applied it
to the OpenTelemetry Demo microservice application. Starting from traces
exported to Jaeger, we automatically constructed a service dependency graph
with Kafka annotations, instantiated all-blocking and async semantics over this
graph, and estimated endpoint availability under fail-stop service failures via
Monte Carlo simulation. A GitHub Actions pipeline ties these steps together and
compares predictions with live chaos experiments on a Docker Compose deployment
of the demo.

Even with its connectivity-only view, the discovered model provides a useful
approximation of endpoint availability across a wide range of failure
fractions. It reproduces the overall degradation curve and correctly reflects
the effect of failure fraction, with modest optimism at low failure rates and
pessimism at very high rates. In this case study, treating Kafka edges as
asynchronous changes predicted availabilities by at most about $10^{-5}$
(0.001 percentage points), indicating that for the exercised endpoints immediate HTTP success is dominated by synchronous dependencies at our experimental resolution.

Future work includes applying the framework to larger benchmarks and industrial
microservices, refining the treatment of asynchronous and partial dependencies
(e.g., broker acknowledgements, queue capacity and eventual processing
requirements), and relaxing the assumption of independent fail-stop faults to
include correlated and gray failures.

\bibliographystyle{splncs04}
\bibliography{refs}

\end{document}